# Chip-Integrated Metasurface Full-Stokes Polarimetric Imaging Sensor

*Jiawei Zuo[1,2], Jing Bai[1,2], Shinhyuk Choi[1,2], Ali Basiri[1,2], Xiahui Chen[1,2], Chao Wang[1,2,3], Yu Yao[1,2]\**

[1]School of Electrical, Computer and Energy Engineering, Arizona State University, Tempe, AZ, USA, 85281

[2]Center for Photonic Innovation, Arizona State University, Tempe, AZ, USA, 85281

[3]Biodesign Center for Molecular Design &Biomimetics

**\*Corresponding author: yuyao@asu.edu**

**ABSTRACT**

Polarimetric imaging has a wide range of applications for uncovering features invisible to human eyes and conventional imaging sensors. Compact, fast, cost-effective and accurate full-Stokes polarimetric imaging sensors are highly desirable in many applications, which, however, remain elusive due to fundamental material limitations. Here we present a Metasurface-based Full-Stokes Polarimetric Imaging sensor (MetaPolarIm) realized by integrating an ultrathin (~600 nm) metasurface polarization filter array (MPFA) onto a visible imaging sensor with CMOS compatible fabrication processes. The MPFA is featured with broadband dielectric-metal hybrid chiral metasurfaces and double-layer nanograting polarizers. This chip-integrated polarimetric imaging sensor enables single-shot full-Stokes imaging (speed limited by the CMOS imager) with the most compact form factor, record high measurement accuracy, dual-color operation (green and red) and a full angle of view up to 40 degrees. MetaPolarIm holds great promise to enable transformative applications in autonomous vision, industry inspection, space exploration, medical imaging and diagnosis.

# INTRODUCTION

The development of imaging systems has profoundly impacted our lives, from smartphone cameras to the most advanced medical imaging equipment and even to space exploration. Besides intensity and color, the polarization state of light, which can change upon emission, scattering, or transmission by an object, is essential for various applications such as target detection [1-3], biomedical diagnostics [4-6], remote sensing [7], defense [8], and astronomy [9], etc. Thus, it is highly desirable to create a compact and economical polarimetric imaging system that not only records the light intensity but also analyzes the polarization state at each pixel to provide more complete information of the target object.

Conventional polarization imaging systems require complex optical components and moving parts, making system miniaturization difficult [10-12]. Moreover, these systems also suffer from reduced frame rates and inaccurate extracted polarization information due to motion in the scene. Monolithic integrated linear polarization imaging sensors have been demonstrated by integrating metallic nanowire-based linear polarization filter arrays onto CMOS imaging sensors [13-16] based on a spatial division measurement approach to avoid moving parts; they are ultra-compact and high-speed, yet cannot perform complete measurement of polarization states due to the lack of materials with strong chirality for ultra-thin circular polarization filters. Various types of circular polarization filters have been studied, including liquid crystal polymers [17, 18], birefringent polymers [19, 20], and micro retarders on metallic nanowires[21]. However, these thin-film structures post various limitations on-chip integration with silicon imaging sensors due to material compatibility issues, environmental stability issues, and potential crosstalk issues due to the relatively thick films required for sufficient polarization filter extinction ratios.

Recent development in optical metasurface and metamaterials has enabled much more compact, flexible, and robust solutions for polarization detection than conventional techniques [22]. Prevailing techniques for metasurface-based polarization imaging can be categorized into dielectric metasurface diffraction grating [23], plasmonic metasurface [24], microscale polarization metalens array [25, 26]. The dielectric metasurface diffraction grating devices [23] are efficient, yet the operation field of view (FOV) is less than 10° and operation bandwidth is limited to 10nm due to dispersive nature of diffraction gratings. The plasmonic metasurface microscale polarization filter arrays [24] exhibited high efficiency and broad bandwidth, but the working wavelength is limited to infrared wavelength due to high plasmonic loss in visible wavelengths. The microscale polarization metalens array exhibits high efficiency and high compactness. Yet the operation bandwidth is limited to less than 10nm due to dispersive nature of metalens in polarization control and focusing. Moreover, the pixel size is limited as optical crosstalk becomes more pronounced as pixel size decreases (7.5% for pixel size of 7 μm and 13% for pixel size of 2.4 μm). Besides polarimetric imagers, many single point polarimetric detectors has also been demonstrated. Among them, various plasmonic structures have achieved broader working wavelengths [27-29], high detection accuracy[28, 30], and direct integration with photodetector[31-33] for infrared wavelengths, yet the employment of plasmonic structures in visible range generally remains challenging due to optically lossy nature of plasmonic structures; dielectric gratings or metalens have achieved ultra-compactness and high efficiency[34, 35] in visible, yet the incidence angle range and operation bandwidth are still small due to their angular and chromatically dispersive nature; metal-dielectric hybrid metasurfaces have been used for chiral metasurfaces and polarization detection with high efficiency and performance for near IR wavelengths [36] and is suitable for extending into visible range. So far, ultra-compact high speed full-Stokes polarimetric imaging sensors for visible wavelengths with high efficiency, high detection accuracy, and broad FOV remains elusive.

Here, we report a chip-integrated single-shot Metasurface-based Full-Stokes CMOS Polarimetric Imaging sensor (MetaPolarIm) with dual-color operation (i.e., red color from 630 to 670 nm and green color from 480 to 520 nm), high measurement accuracy and an FOV up to 40°. The MetaPolarIm is composed of a subwavelength-thick hybrid metasurface polarization filter array (MPFA) integrated on a conventional CMOS imaging sensor. The device design adopted the spatial division measurement approach [37-40] to obtain full Stokes polarimetric images at one snapshot with imaging speed ultimately limited by the CMOS imaging sensor. We achieved dual-color operation for the circular metasurface polarization filter utilizing the material dispersion and structural engineering of Si nanograting metasurfaces. The device design and fabrication process are compatible with CMOS technology. The measurement errors for all Stoke parameters are <2% for red and green at normal incidence and remain <4% over the FOV up to 40 degrees. Compared with state-of-the-art technologies[11, 12, 16, 17, 23-26, 41], the advantages of our device are ultra-compactness, compatibility with CMOS technology, single-shot operation, dual-color operation, high FOV and high measurement accuracy (see table S1 and S2 in Supplementary Information for a detailed comparison with the state-of-art polarimetric imaging sensors).

## RESULTS AND DISCUSSIONS

### DESIGN CONCEPT

The chip integrated MetaPolarIm (Figure 1a) comprises an MPFA and a commercial CMOS imaging sensor beneath it, as shown in Fig. 1b, and is almost the same size as the conventional CMOS imaging sensor. The MPFA consists of over 75,000 microscale polarization filters. Each super-pixel has two pairs of circular polarization (CP) filters ($P_5$ and $P_6$), and four linear polarization (LP) filters ($P_1$ to $P_4$). Each polarization filter is defined as one sub-pixel and integrated on top of one or a few imaging pixels of the CMOS imaging sensor underneath (Figure 1c). The polarization state at each super-pixel can be obtained by the measurement results of a combination of LP and CP sub-pixels and their corresponding instrument matrix[28], as discussed in detail in Materials and Methods. Figure 1d shows the schematics of the CP and LP filter designs, which are based on metal-dielectric hybrid metasurfaces. Both LP and CP polarization filters have a thickness of less than a wavelength, resulting in a highly compact form factor for the demonstrated full-Stokes polarimetric imaging camera.

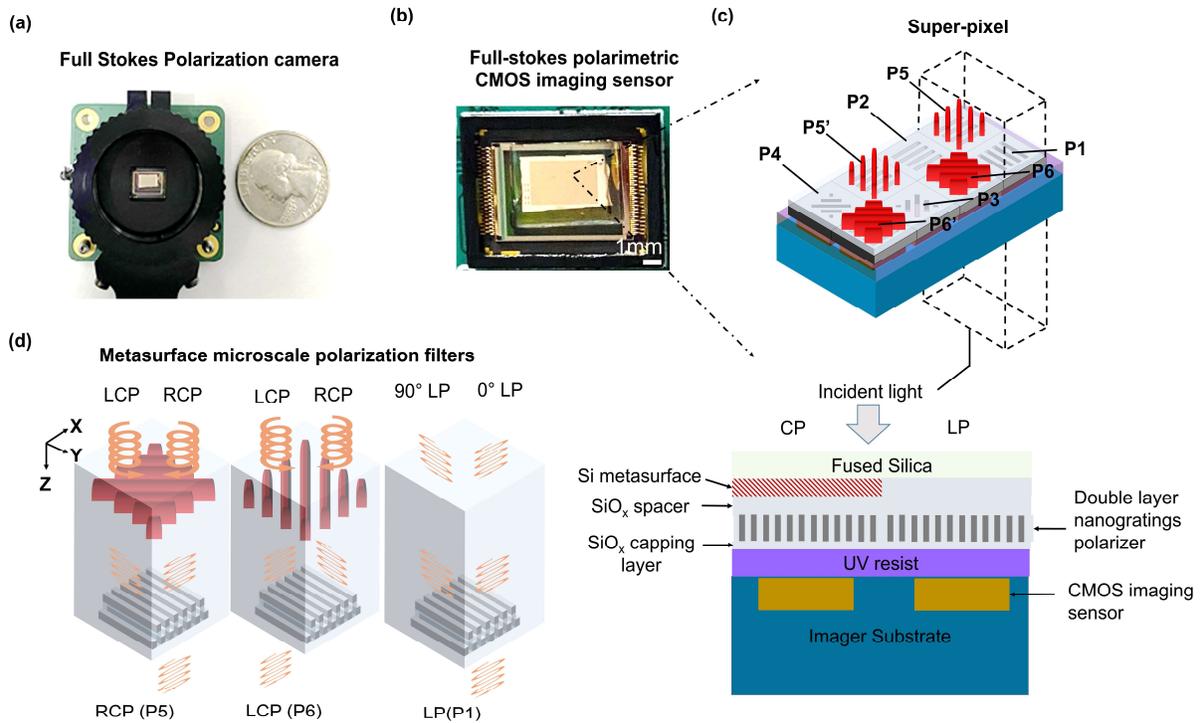

*Figure 1. CMOS integrated full Stokes polarimetric imager with dual operation wavelength. (a). Image of full Stokes polarization camera beside a U.S. dollar coin (lens not attached). (b). Image of full Stokes polarimetric CMOS imaging sensor (c) Top: 3D Conceptual illustration of chip integrated full Stokes CMOS polarimetric imaging sensor. Here P1-P4 denotes the LP filters*

*with transmission axes at 0°, 90°, 45°, 135° respectively. P5,P5' and P6,P6' denote chiral metasurface filters transmitting right-handed circularly polarized (RCP) and left-handed circularly polarized (LCP), respectively. Here, P5 and P5', P6 and P6' are identical in dimensions respectively. bottom: 2D cross-section of the chip-integrated polarization imaging sensor. (d) 3D conceptual illustration of a pair of chiral metasurfaces responsible transmitting RCP and LCP light, respectively (P5, P6) and a LP filter (P6)*

The metasurface-based microscale polarization filters were all made of CMOS compatible materials, i.e., Aluminum (Al), silicon and silicon oxide. We chose vertically coupled double-layered gratings (VCDGs) made of Al at subwavelength scale (Figure 2a) as the LP filter due to its high linear polarization extinction ratio (LPER) and fabrication simplicity. For incident light with electric field vector oriented in parallel with the nanogratings (i.e., y-axis in Fig. 2a), it is almost completely reflected by the double layers of gratings. For incident light with electric field vector oriented vertically to the nanogratings (x-axis in Fig. 2a), it can couple effectively into the gap plasmon modes in the deep subwavelength-scale vertical nanogap ($g<\lambda_0/10$) between the top and bottom Al nanowires and then transmit through the grating layer with high efficiency, as shown in Fig. 2b. The small vertical gap between the two coupled grating layer is essential for high transmission and LPER. We have optimized the design parameters of the VCDG, including the vertical nanogap, grating duty cycle, period and Al thickness to maximize the LPER and transmission efficiency over the visible wavelength range (Figure S1). Our designed VCDG has LPER over 1000 in the visible range with maximum efficiency of 56% and 47% at 650nm and 510nm, respectively, as shown in Figure 2c. Compared to single-layered metallic grating linear polarizers of the same grating thickness [13], the VCDGs has 2 order of magnitude higher LPER and slightly lower transmission efficiency for the transmitted polarization (Figure S2). Moreover, the fabrication process of VCDGs is much simpler and has higher success rate, since it does not require etching and lift-off of Al metal films with thickness of a few hundred nanometers, which are very challenging for nanofabrication of feature size down to tens of nanometers.

We designed the CP filters based on metal-dielectric hybrid chiral metasurfaces by integrating a metasurface quarter waveplate (QWP) onto Al VCDGs. The basic concept is similar to our previous work where high-efficiency CP filters were demonstrated for NIR wavelength [36]. Yet, instead of using an array of Si nanopillars for the metasurface QWP in visible wavelengths, we engineered chiral metasurface structures (Figure 2d) utilizing silicon nanogratings with much more feasible dimensions for nanofabrication and broadband operation wavelengths. Compared with nanopillars, silicon nanogratings exhibit a larger artificial birefringence $\Delta n$ ($\Delta n = n_V - n_U$, where $n_V$, $n_U$ are defined as the effective refractive index along nanograting direction and its orthogonal direction), therefore requiring a much smaller height-to-width aspect ratio (~1.3 according to Figure S3) for the same $\frac{\pi}{2}$ phase difference engineering than that of Si nanopillar design (~4 according to Figure S4). The large artificial birefringence of Si nanogratings stems from the anisotropic near-field distribution under different incident light polarization (illustrated in Figure 2e, a cross-sectional view in Figure S5). When incident light is polarized along the U axis (fast axis), the Si gratings mainly support leaky modes [42-44], and the electric field intensity is highly localized in $SiO_x$ gaps between Si (Figure S5a). On the other hand, for incident light polarized along the V axis (slow axis), the Si nanogratings support modes with electric field mostly located inside Si (Figure S5b). Therefore, the effective refractive index along U axis is lower than that along V axis. The CP filters were realized by integrating an aluminum VCDG linear polarizer with its optical axis oriented at +/-45 degrees (for LCP/RCP filters) to that of the silicon nanogratings. Figure 2d illustrates the working principles. In the LCP filter design, the LCP incident light is converted into LP light oriented along x axis by the Si nanogratings and then passes through the VCDGs with high transmission, while the RCP incident light is converted into LP light oriented along y axis and blocked by the VCDGs, resulting in very low transmission (<1%). Moreover, we also demonstrated that by engineering the dispersion of silicon nanogratings, one can achieve CP filters simultaneously for two wavelength ranges. By properly adjusting the silicon nanograting period, duty cycle and thickness, we achieved a phase difference between fast and slow axes to be $\frac{\pi}{2}$ at red color (629 nm, Figure S6). Meanwhile, Si nanogratings acts as QWP at a shorter wavelength centered around 500nm (green color), mainly due to its highly dispersive near-field distribution (Figure 2g, cross-sectional near-field distribution in Figure S7), and therefore produce an advanced phase $\frac{3}{2}\pi$ along the fast axis (Figure S6). In this case, LCP incident light around 500 nm is converted into LP light polarized along y axis and then blocked by VCDGs, while RCP incident light is converted into LP light polarized along x axis and transmits through the VCDGs (Figure. 2d). Figure 2f shows the transmission spectra (left axis) and the circular polarization extinction ratio (CPER, right axis) of an LCP filter for red color 550~750nm, obtained by full wave simulation. The device has a maximum CPER of over 600 with a transmission efficiency of 61.5% at 650nm. Moreover, the device design can operate from 600nm to 700nm with CPER over 10. For shorter wavelength around 500 nm, the chiral metasurface detects opposite CP handedness at a shorter wavelength, as shown in Figure 2h. The chiral metasurface exhibits a maximum CPER of 170 with a transmission efficiency of 24.5% at 510nm and 450nm to 525nm with CPER over 10. The dual working wavelength of Si nanograting broadened the operation wavelength range with a total bandwidth of 175nm(CPER>10), which is one order higher than other grating diffraction-based designs[23, 25].

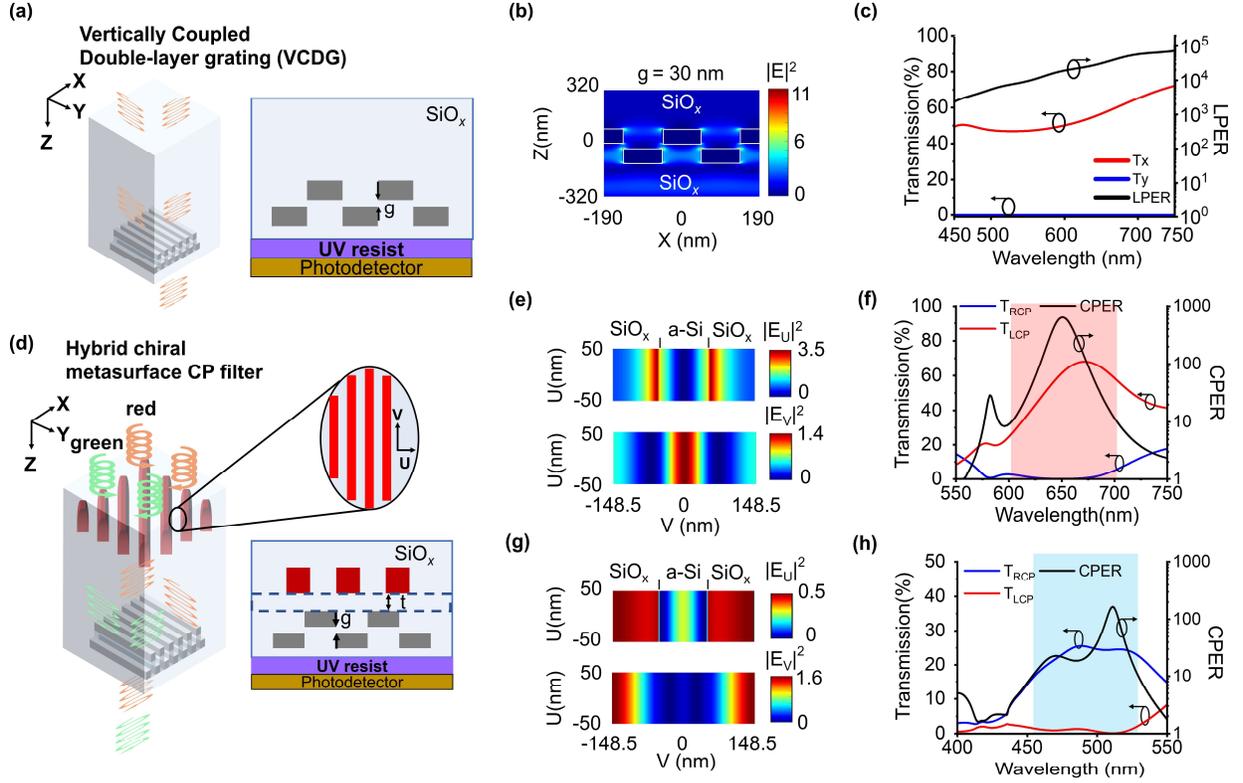

*Figure 2. Design of the chiral metasurface and VCDG for CP and LP detection.* *(a). 3D schematic and 2D cross-section of VCDG. The thickness of Aluminum (Al), width, period, and the vertical gap of VCDG are $t_{Al}$= 80nm, $p_1$=190nm, $w_{Al}$ = 95nm, and g=30nm, respectively. (b). Cross-sectional view of VCDG near field distribution with input light (650nm) polarized along x axis. (c). Transmission spectra and LPER of VCDG with input light polarized along x and y axis respectively. (d). 3D schematic and 2D cross-sectional view of the top layer Si nanograting and bottom layer VCDG, respectively. The thickness, period, width, and tilted angle of Si nanograting are $t_{si}$=130nm, $p_{Si}$=297nm, and $w_{Si}$=100nm. The thickness of Aluminum (Al), period, and vertical gap of bottom layer VCDG are $t_{Al}$= 80nm, $p_2$= 210nm, and g=30nm, respectively. The thickness of the $SiO_x$ spacer layer is t=400nm. (e). Near field distribution of the Si grating when incident polarization is along the width of the Si nanogratings (U axis) and length of the Si nanogratings (V axis) at 629nm. (f). Simulated transmission spectra(left) and CPER (right) of dielectric-metal*

*hybrid chiral metasurface at 550nm-750nm. (g). Near field distribution of the Si grating when incident polarization is along the U and V axes at 500nm. (h) Simulated transmission spectra(left) and CPER (right) of dielectric-metal hybrid chiral metasurface at 400nm-550nm.*

**DEVICE FABRICATION AND CHARACTERIZATION**

Figure 3a shows the fabrication process flow of the MPFA. First, amorphous silicon (*a*-Si) and SiO$_x$ were deposited onto fused silica wafers by plasma-enhanced chemical vapor deposition (PECVD), followed by electron-beam lithography (EBL), lift-off, reactive-ion etching (RIE) of SiO$_x$ mask, and inductively coupled plasma etching (ICP) of *a*-Si to form Si nanogratings (SEM image of fabricated Si nanogratings are included in Supplementary Information, Figure S8). Then, a dielectric spacer layer (520nm) of SiO$_x$ was sputtered onto Si nanogratings, followed by EBL patterning and Cr deposition for the VCDG layer. After lift-off, the formed Cr nanogratings were used as a hard mask to transfer the grating pattern into the SiO$_x$ spacer layer (depth 110nm) by reactive ion etching (RIE). Then 80nm Aluminum (Al) was deposited by electron beam evaporation to form the VCDG structure. Finally, the fabricated MPFA was covered with 200nm-thick sputtered SiO$_x$ as a protection layer. The fabricated MPFA sample was cut into 4mm*3mm pieces using a wafer dicing saw and bonded onto a CMOS imaging sensor (Sony IMX 477) using a home-built UV bonding setup. The schematic of the UV bonding setup is included in Supplementary Information (Figure S9), and more detailed information about the fabrication process and UV bonding process are discussed in the Materials and methods section. The fabricated MPFA (Figure 3b) consists of over 129K microscale metasurface polarization filters (MMPF), with a total area of 3.65 x2.43mm$^2$. The scanning electron microscope (SEM) images of one super pixel (center) and 8 different MMPFs, i.e., subpixels, are shown in Figure 3c. We used metal frames around each MMPF to reduce the optical cross talk between the adjacent pixels to less than 1%. Here the metal frame (width 4.65μm) is chosen to minimize the amount of oblique incidence light, which can have a serious impact given a thick spacer layer (around 700nm, including bonding resist and SiO$_x$ capping layer) between the MPFA and CMOS imaging sensor and absence of micro lens array on the metasurface array. Yet, the metal frame width could be further reduced by directly fabricating the MPFA on Si photo diode and/or adding micro lens array to minimize cross-talk between adjacent pixels.

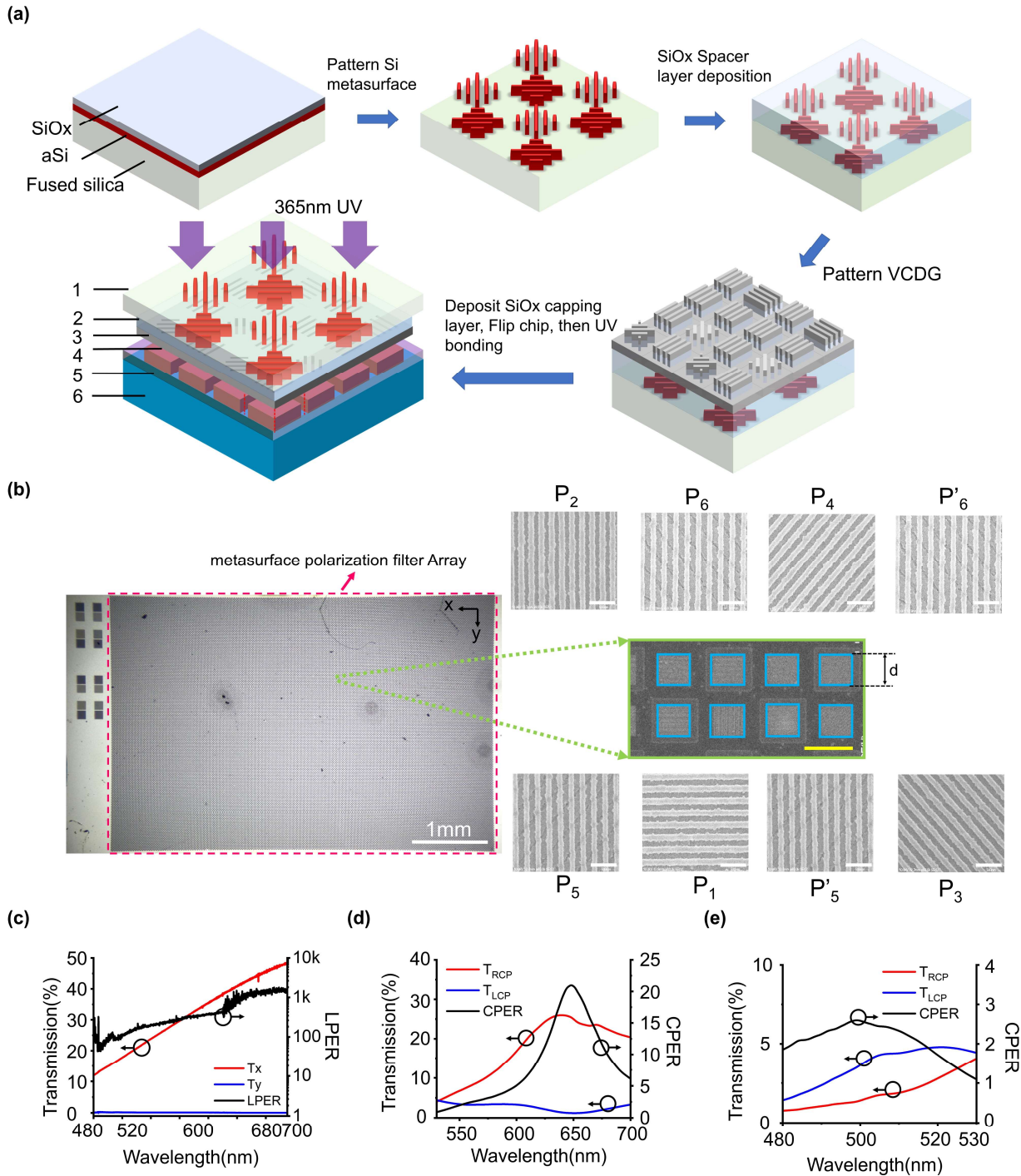

***Figure 3. Device fabrication and characterization.*** *(a) Schematic of the device fabrication process. 1: fused silica wafer, 2: SiO$_x$ spacer layer, 3: VCDG, 4: SiO$_x$ capping layer, 5: spin coated ultra-violet (UV) resist onto the CMOS imaging sensor, 6: CMOS imaging sensor (b) The microscopic photograph(left) and the SEM image of one super pixel among the fabricated MPFA*

*(right), scale bar:10μm. 90°,45°,0°,135° LCP, and RCP on each sub-figure indicate the polarization state each metasurface filter transmits. Scale bar: 500nm (white). The size of CP and LP micro-filters are 6.2μm by 6.2μm, with spacing ... between adjacent filters to minimize crosstalk (c,d) Measured transmission and CPER of the chiral metasurface at 480nm-530nm and 530-700nm, respectively. Legend RCP, LCP indicates the input CP handedness. (e). Measured transmission and LPER of fabricated VCDG under TM and TE polarization.*

The performance of fabricated MPFA was first characterized by a visible spectrometer with a broadband LP polarizer with LPER over 1000 and a broadband QWP. (See more details of the setup in the Materials and methods section and Figure S10 in Supplementary Information). Figure 3c shows the measured transmission spectra of the VCDG (oriented at 0°) patterned on the $SiO_x$ spacer layer without the Si metasurface buried underneath, which shows an efficiency of 35% and LPER over 400 around 650nm. Noticeably the VCDG provides LPERs from 100~400 from 520nm to 700nm, offering broadband LP detection with high accuracy, which is 1 order higher than single layered gratings presented in literature[13]. Compared with simulation results, the transmission efficiency and LPER of fabricated VCDG are still lower. We attributed such performance degradation to the surface roughness of sputtered $SiO_x$ spacer layer (Ra=8.43nm, as shown in Figure S11) as well as the surface and edge roughness introduced in the metal evaporation. As a comparison, we fabricated VCDG on a fused silica wafer with similar dimensions by EBL patterning followed by Al E-beam evaporation. We increased the Al deposition rate and vacuum level to improve Al film quality. The results showed reduced surface and edge roughness of the fabricated VCDG (Figure S12a). The measured transmission efficiency was also improved by almost twice across the visible wavelengths (results also shown in Figure S12). Thus, we expect to improve the VCDG linear polarizer performance by improve the quality of the SiO$_x$ spacer and Al thin film.

Figure 3d shows the measured circular dichroism (CD) spectra of a CP filter that transmits RCP light while rejects LCP light at wavelengths around 650 nm. This CP filter provides CPERs of more than ten over a wavelength range of 60nm, about 10% of the optimal operation wavelength at 650nm. Figure 3e shows the measured circular dichroism (CD) spectra of the same device at shorter wavelengths around 500 nm. As expected in simulation, the CP filter transmits LCP light while rejects RCP light with a CPER of ~2.5 at 500nm. Compared with simulation results, the chiral metasurface-based CP filters showed reduced optical performance. Besides the surface and edge roughness that led to lowered efficiency and LEPR or the VCDG structure, the Si nanogratings buried under the spacer layer leave an increased surface roughness (Ra = 28.3 nm, as shown in Figure S13) of the SiO$_x$ spacer layer, resulting in a significant decrease in LPER and thereby reducing the CPER of the chiral metasurface. We measured the LPER (Figure S16) of the bottom layer VCDG of a chiral metasurface (device B) with similar dimensions (Figure S14). The LPER of the bottom layer VCDG is only ~15. As a result, the CPER of device B is reduced to below 10 (Figure S15). All the results and analysis show that surface planarization of the dielectric spacer layer is required for the high optical performance of such a bi-layer dielectric-metal hybrid chiral metasurface design. We plan to address surface planarization issues for future integration of multi-layer metasurface devices.

## HIGH-ACCURACY POLARIMETRIC DETECTION

The simplest method for obtaining the Stokes parameters in spatial division method are directly subtracting the intensity of 0°,90°,45°,135° LP, and RCP and LCP components [45]. Yet the accuracy is limited by the LPER and CPER of the polarization filters and their wavelength dependence [46]. Here, we first performed calibration to obtain the instrument matrix $A$ of the polarimetric imaging sensor. Then the Stokes parameters $S$ can be obtained using the linear equation: $S = A^{-1}I$, where $I$ denote the instrument matrix of the polarimetric imaging sensor. (See Supplementary Information section 6 for more details about the Instrument matrix calibration process). This method can greatly increase polarization detection accuracy [28, 47].

Figure 4a shows a customized optical setup for calibrating the device. A color-filtered, uniform, collimated beam with sufficient spot size is incident onto the full Stokes polarization imaging sensor, which is mounted onto a rotational stage to control the light incidence angle $\varphi$. During the calibration, $\varphi$ is kept at 0° (normal incidence) to obtain the instrument matrix for each microscale polarization filter. We apply a polarization state generator (polarizer and QWP) to generate arbitrary polarization states. With each polarization state input, a snapshot of the transmitted light intensity through MPFA was taken to obtain the intensity vector [47]. The instrument matrix $A$ of MPFA could be readily obtained after sufficient polarization state input to form an intensity matrix. Besides, a spatially over sampling approach was applied during the instrument matrix calculation to enhance the imaging resolution (see Figure S17 in Supplementary Information for more details). After calibration, we performed full-Stokes polarization detection on the same setup at different incidence angles, using the instrument matrix $A$ calibrated at $\varphi = 0°$. Note that we did not perform the instrument matrix calibration at different incident angles because in most imaging applications, the imaging sensor will receive light from different incident angles all at once and usually could not distinguish light from different angles.

We first generate 18 arbitrary polarization states and measure them using a conventional PSA (see Materials and Methods), then the generated polarization states are incident onto the device uniformly. Figure 4 b,c depicts the measured Stokes parameter $S_i^j$ and their reference values $S_{R_i}^j$ measured by polarization state analyzer (PSA) ($i=1,2,3; j=$ A,B… R) on the Poincaré sphere under red and green color input. In total, we chose 18 reference polarization states sparsely distributed in all eight quadrants of the Poincaré sphere to verify the full Stokes polarimetric detection accuracy better. Here $S_i^j$ is averaged out over the entire MPFA : $S_i^j = \frac{\sum_{a=1,b=1}^{n,p} S_{i\,a,b}^j / S_{0\,a,b}^j}{n \times P}$, i=1, 2, 3, j= 1,2… 18, n=335, p=221. where $S_{i\,a,b}^j / S_{0\,a,b}^j$ represents normalized Stokes parameters measured by each pixel. Figure 4d shows the measurement error $\Delta S_i^j$ for each polarization states at $\varphi = 0°$ under red and green colors. Here $\Delta S_i^j$ is defined as: $\Delta S_i^j = S_i^j - S_{R_i}^j$ (i=1, 2, 3, j=1,2… 18). The mean absolute error (MAE) $\frac{\sum_{j=1,}^{18} |\Delta S_i^j|}{18}$ ($i$=1,2,3) for $S_1$, $S_2$, $S_3$ are 1.84 %, 1.93 %, 1.79% for green and 1.03%, 1.43% , 1.99% for red, respectively. Measured $\Delta S_i^j$ of other incidence angles is included in Supplementary Information (Figure S18 for red and Figure S19 for green color). Based on the Stokes parameters measured, we calculated the angle of polarization (AOP=$\frac{1}{2}\arctan\frac{S_2}{S_1}$), degree of circular polarization (DOCP=$S_3/S_0$), and degree of linear polarization (DOLP=$\sqrt{S_1^2 + S_2^2}/S_0$). Here $\sigma_i$ is written as: $\sigma_i = \frac{\sum_j^{18} \sigma_i^j}{18}$ i=1,2,3, j=1,2…18 where $\sigma_i^j$ is defined as the standard deviation of Stokes parameter measurement error of MPFA, denoted as: $\sigma_i^j = \sqrt{\dfrac{\sum_{a=1,b=1}^{n,p}\left(S_{i\,a,b}^j/S_{0\,a,b}^j - S_{R_i}^j\right)^2}{n\times p}}$. Measured MAE and averaged standard deviation $\sigma_i$ for $S_1$, $S_2$, $S_3$, AOP, DOLP, and DOCP at different incidence angles is included in Supplementary Information (table S3 for red color and table S4 for green color). MAE for AOP, DOLP, and DOCP are 0.26°, 1.41%, 1.99% for red and 0.78°, 1.26%, 1.79% for green at $\varphi = 0°$. When $\varphi$ is within ±20°, MAE for $S_1$, $S_2$, $S_3$ can maintain less than 4% for red. At ±30° incidence, MAE for $S_3$ increases to 17.51%, this is because CPER of chiral metasurface reduces by two orders compared to normal incidence, indicated by FDTD simulation (Figure S20a). Similarly, the device has an MAE of less than 4.1% for green color input with incidence angle within ±5°, with incidence angle up to ±10°, the detection error of S3 increases to 15% due to a decrease of CPER (Figure S20b).

We then evaluated the uniformity of detection accuracy across the polarimetric imaging sensor. Figure 4e shows the distribution of measurement error of AOP, DOCP, and DOLP for polarization state D at different incidence angles. Over 90% of the polarimetric imaging pixels provide small measurement errors for DOLP (<2%), DOCP (<2%) and AOP (0.5°), respectively, for both red and green color at normal incidence. A larger AOP, DOCP and DOLP error range can be seen when incidence angle increases. When $\varphi$ is within ±20°, over 80% …. are less than 10%, 10%, 2°. The increase of the non-uniformity of measurement errors at oblique incidence is majorly due to instrument matrix is calibrated at $\varphi=0°$ (normal incidence), causing calibration errors when the imaging sensor is measuring polarization states in oblique incidence as MPFA has different CPER upon oblique incidence (Figure S20). Measurement error distribution of other polarization states under the red/green input can be found in Supplementary Information session 7 (Figure S21 to Figure S38).

So far, we have shown that with the instrument matrix method, our full Stokes polarimetric imager allows highly accurate polarization state detection at red and green colors with a single snapshot. Moreover, we have shown that the instrument matrix method can be applied to an array of micro polarization filters with a standard deviation of less than 1% for both red and green color under normal incidence and are less than 5% for red color within ±20° oblique incidence. Our polarimetric imager, together with the instrument matrix reconstruction method, has the potential to achieve high accuracy full Stokes parameter imaging with dual-wavelength coverage.

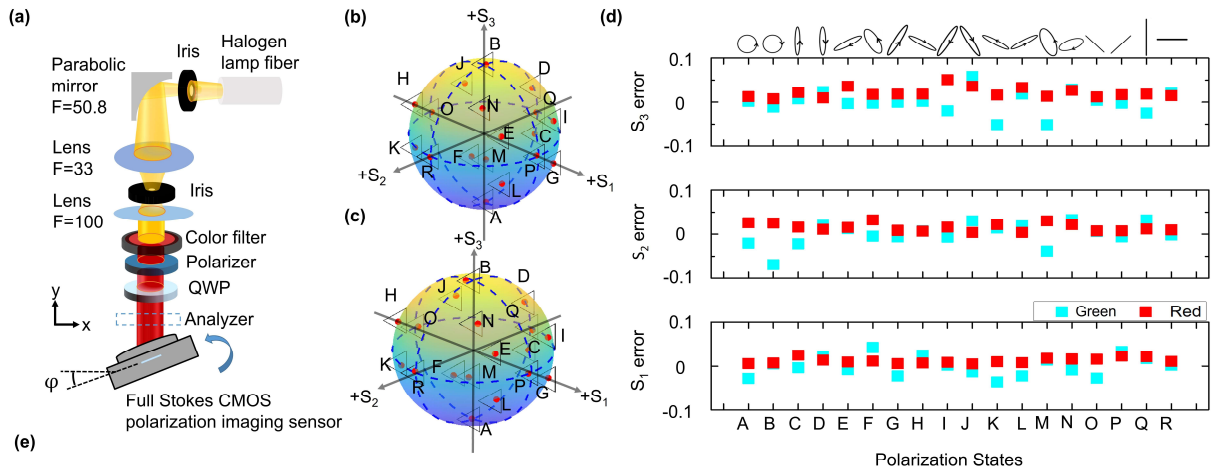

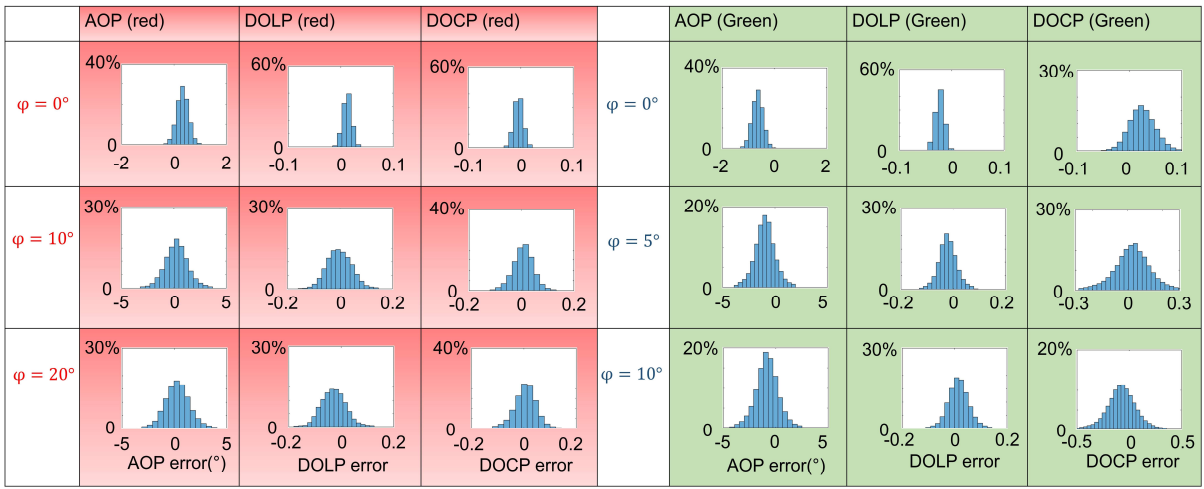

*Figure 4. High Accuracy full Stokes polarization detection.* *(a) A schematic of the customized experimental setup for generating arbitrary polarization states for full Stokes polarization*

*detection, φ denotes the camera rotation in azimuth angle for oblique incidence measurement. Here parabolic mirror, 2 iris and 2 lens were applied to generate collimated beam with divergence angle less than 0.5°, unit: mm. (b,c) Illustration of measured polarization state (Red dot) and its input reference (Triangle) distribution on Poincaré sphere in red and green color, respectively. (d) Error distribution for Red/Green color full Stokes parameter detection measurement result. (e) AOP, DOLP, and DOCP detection error distributions of all MPFA pixels for polarization state D at normal incidence and oblique incidence of red color (left) and green color (right). In all the plots, x-axes represent the errors and y axes represent the corresponding percentage of pixels.*

#### DUAL-WAVELENGTH FULL-STOKES PARAMETER IMAGING

Here we demonstrate the full Stokes polarization imagery with several objects for proof of concept. The experimental setup for full Stokes polarization imaging is shown in Supplementary Information (Figure S39). A 40-nm bandpass filter centered at 650nm and 500nm was applied separately in front of the mercury lamp fiber. We took polarization images of several real-life objects carrying the polarization information. The objects were positioned behind the paper allowing diffused light to transmit from behind. Each image section includes an object photo taken by a cell phone camera; an image with colorful background identifying the color; the raw exposure $S_0$; the angle of polarization (AOP); the degree of linear polarization (DOLP); the degree of circular polarization (DOCP) and the degree of polarization (DOP). Next, we discuss those sections respectively. Section A shows a pair of 3D glasses consisting of opposite CP information. The handedness of the input CP cannot be seen in the sample photo and intensity image but is clearly shown in the DOCP image. We notice that the values of DOCP and AOP of the right glasses are different when taken with red and green colors, indicating differences in the transmitted polarization state of 3D glasses under a different color. This example reveals potentials of MetaPolarIm in virtual reality and augmented reality technology where CP glasses are widely used[48]. Section B shows a pair of plastic goggles. In the sample photo, the plastic goggle looks transparent. However, the DOCP image of goggles looks rather un-uniform because of the birefringence of plastic stemming from stress. In addition, the goggles' DOCP image under the red and green color light input shows readily different distribution, indicating plastics' birefringence dependence on the input color. This example clearly exemplifies the advantage of full Stokes polarization imaging under dual operation wavelengths, which could be applied to numerous applications such as industry imaging and remote sensing. Section C examines the polarization information of sunglasses. The red and green DOLP images and DOP images both show high values, while the DOCP value of the glasses region is nearly 0, indicating the unpolarized light gets linearly polarized upon transmission through the sunglasses. Both red color and green color images show similar conclusions, revealing the broadband linearly polarized characteristics of the sunglasses. This example reveals potential of MetaPolarIm in applications such as glare reduction and contrast enhancing of objects with polarization information. Section D depicts a plastic cage imaged with 0°LP as the input background. High DOCP values and un-uniform AOP image in the cage area are due to the inner stress of the plastic material upon molding, giving rise to birefringent material optical characteristics. We notice that the red and green color DOCP are readily different, which reveals differences in the inner stress distribution of the plastic cage under different colors. This example reveals potential of MetaPolarIm in applications in material stress detection. Section E examines a simple test; the LCP camera filter is circularly polarized; this is not visible to the traditional imaging sensor but is clearly shown in red and green color DOCP images. Besides, thanks to dual operation wavelength, our sensor shows that the AOP of the CP filter is readily different in red and green color, indicating the light transmitted through the LCP filter with different wavelengths shows different polarization states.

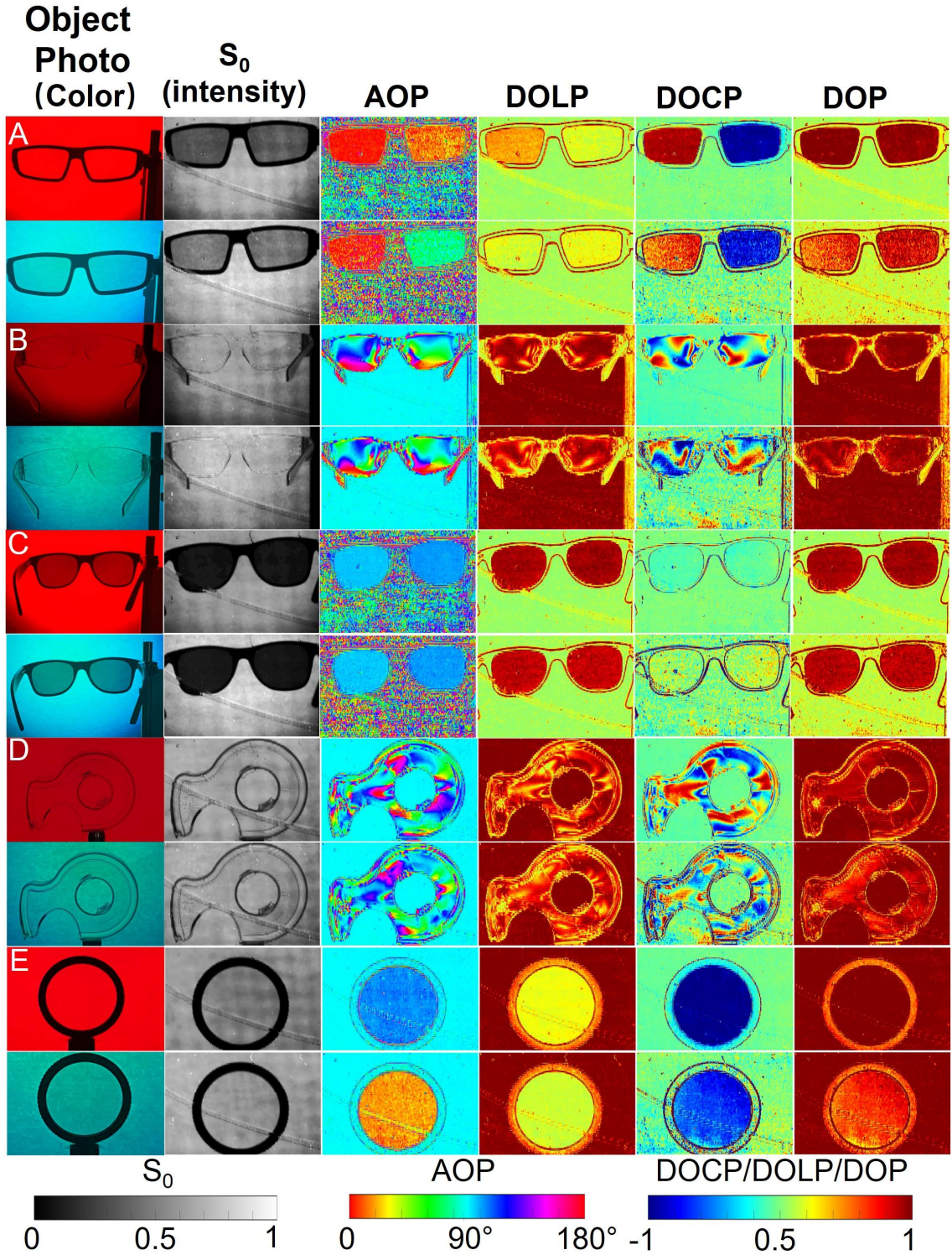

***Figure 5. Full Stokes polarization images of various objects.*** *Full Stokes polarization imaging of objects carrying polarization information. Section A:3D glasses with the unpolarized background; section B: Plastic goggles with CP input as background. Section C: Sunglasses with unpolarized input as background. Section D: Plastic cage, with CP input as background. Section E: A camera CP filter, with CP input as background. In each case, an object photo was taken by a cell phone camera. "Color" image indicates Bandpass filters applied, 630nm-670nm (Red) and 480nm-520nm (Green) in front of the light source.*

## CONCLUSION

In this work, we designed and fabricated metasurface-based microscale polarization filter array composed of broadband microscale linear polarization filters and dual-color (green and red) circular polarization filters (56K pixels). We then integrated the MPFA onto the CMOS sensor and calibrated the sensor polarization detection with the instrument matrix calibration method. With calibration, we achieved high Stokes measurement accuracy: Averaged measurement Error less than 2% for $S_1$, $S_2$ $S_3$ in red and green color. Moreover, our polarization imaging sensor can maintain an error of less than 5% up to ±20° oblique incidence for red color and ±5° for green color. Finally, we demonstrated the full Stokes polarization imaging in real-life objects invisible to the traditional imaging sensor at red and green color with total operation bandwidth of 80nm. From the polarization images of objects, we find polarization information carried by these objects is color-dependent, revealing the advantage of dual-wavelength operation. Overall, our full Stokes polarization sensor is a mini, ultra-compact, full Stokes parameter imaging device that could be widely adopted in various real-life applications, such as enhancing contrast in machine vision, material index sensing, and biomedical imaging.

## Materials and methods

**Simulations**

FDTD simulations from Lumerical Inc. FDTD solver were applied to calculate the transmission efficiency, CPER of the chiral metasurface, and LPER of double-layer gratings. The real optical material refractive index of *a*-Si and Aluminum measured by UV-NIR spectroscopic ellipsometry (J.A. Woollam, M-2000) was applied to FDTD material explorer to calculate the device performance precisely. Specifically, for the simulation of double-layer gratings, the plane wave along the grating width and grating length direction was applied to calculate the LPER and efficiency of double-layer gratings. For the chiral metasurface, a tilt angle of 6° of the Si grating obtained from SEM images was considered in the simulations. Two orthogonally linearly polarized plane waves with a phase difference of $\pm \pi / 2$ were super positioned to represent LCP/RCP light input, respectively. In all FDTD simulations, we simulate one unit cell and apply periodic boundary conditions along the in-plane direction. The simulation convergence auto shut-off level was set to $10^{-5}$. The mesh size was set to 2nm for higher simulation accuracy. For oblique incidence, we use the BFAST plane wave type to maintain the oblique incidence angle the same for all wavelengths.

**Fabrication**

1) Si nanograting: Fused silica wafer was cleaned by RCA-1 cleaning, then amorphous silicon (α-Si) of 130 nm was deposited by plasma-enhanced chemical vapor deposition (PECVD) (Oxford PlasmaLab 100, 350°C/ 15W) on fused silica wafer, followed by deposition of 60 nm SiO$_x$ (350°C/ 20W) without breaking vacuum as a hard mask layer. 10 nm Cr layer was then deposited by thermally evaporating (Denton benchtop turbo) as the discharge layer during 1$^{st}$

EBL exposure. Double-layer polymethyl methacrylate (PMMA) resists (70 nm 2.5% 495k followed with 50 nm 2% 950k) were coated, followed by 2-minute post-baking at 180℃. A pattern composed of 168 by 56 super-pixels array was written with a JEOL JBX-6000FS EBL machine working at 50keV with a current of 500pA. After exposure, the sample was developed for 2 minutes. The developer is a mixture of methyl isobutyl ketone (MIBK) and isopropanol (IPA) with a mixing ratio of 1:3. Next, the sample is cleaned with 30 seconds of Oxygen plasma (PIE Scientific Plasma cleaner, immersion mode $O_2$ 10sccm /20W) to remove PMMA residue on the exposure region. Next, 3nm Cr adhesive layer and 12nm $SiO_x$ were deposited by electron beam evaporating (lesker #3). Cr and $SiO_x$ were lifted off by soaking in warm acetone (60℃) for more than 12 hours, followed by acetone gun cleaning. After the lift-off process, a $SiO_x$ nanostructures array was formed, which masked Cr discharge layer etching by Reactive Ion Etching (RIE) (OXFORD PLASMALAB 80PLUS, Cl2/O2: 9/3 sccm, 10mTorr, D.C. bias/power 18V/70W). An isolated Cr/ $SiO_x$ layered nanostructure mask were thus formed, which then masked anisotropic etching of 60nm $SiO_x$ hard mask by RIE (Plasma-Therm RIE 790, $CHF_3$/ $O_2$ 40/3 sccm, 40 mTorr, 250 W). The dry etching of $SiO_x$ stopped at the α-Si layer; during the dry etching procedure, 12nm $SiO_x$ on $SiO_x$/Cr layered structure was consumed. Then the Cr was removed by CR-4s (Greentek) etchant, and the α-Si layer was etched by ICP-RIE (ICP/bias power of 250/140 W, 10 mTorr, $Cl_2$: Ar=100/5 sccm) using the SiOx mask to complete Si nanograting fabrication.

2) Spacer deposition: The samples were brought into the sputtering chamber (Lesker PVD 75) and covered with a 520 nm $SiO_x$ spacer layer (250W) at a rate of 0.6 Å/s.

3) Vertically coupled double-layered Al gratings: After spacer layer deposition, double-layer PMMA (70 nm 2.5% 495k followed with 50 nm 2% 950k) were coated again, post-baked, and exposed by EBL aligned to the first layer. Then the sample was cleaned with Oxygen plasma to remove residual PMMA on the exposed region. Next, 3nm Cr and 12nm $SiO_x$ were deposited and lifted off, as mentioned above, to form a $SiO_x$ mask for Cr discharge layer etching. Next, Cr discharging layer is etched by RIE to form Cr/ $SiO_x$ layered nanostructures, which masked 100nm $SiO_x$ RIE etching to form $SiO_x$ nano-gratings. Then 2nm Cr and 80nm Aluminum is deposited by E-beam evaporation, forming vertically coupled Aluminum (Al) gratings.

4) U.V. bonding: After sample fabrication is completed and essential device characterization, the sample was then cut into 4mm*3mm by a dicing saw. A CMOS sensor IMX477 was customized to remove the cover glass, micro lens, and Bayer pattern by MaxMax. corp ltd. Then it was spin-coated with 90% UV at a spin speed of 3000rpm/S; the sample was then visually aligned and bonded onto the CMOS sensor by a homemade transfer setup. The detailed schematic of homemade transfer setup is illustrated in Supplementary Information (Figure S9). Afterward, U.V. resists were cured by a 365nm U.V. lamp(100W) for 20 minutes of illumination.

**Measurement**

*Device Transmission and Extinction Ratio Characterization.* For the chiral metasurface, the unpolarized laser was first polarized by linear polarizer (WP25M-UB by Thorlabs, Inc.) and super achromatic QWP (SAQWP05M-700 by Thorlabs, Inc.) to generate LCP, RCP input, respectively. The CP light is then focused onto the sample with a focal spot size of 15um in diameter. The transmission efficiency was then measured using Olympus BX53 fluorescent microscope and Horiba iHR320 visible spectrometer. The CPER of LCP chiral metasurface was calculated using the formula: $E = T_{LCP}/T_{RCP}$, where $T_{LCP}$, $T_{RCP}$ denotes the transmission efficiency of LCP and RCP input, respectively. As for double-layer gratings, the transmission efficiency of TE-mode TM mode linearly polarized input was measured respectively to calculate LPER, using the equation LPER=$T_x$/$T_y$, where $T_x$, $T_y$ denotes the transmission efficiency of LP input along x axis and y axis, respectively.

*Instrument matrix calibration for polarimetric Imager.* Eight polarization states were induced with a broadband linear polarizer (WP25M-UB by Thorlabs, Inc.) and super achromatic QWP (SAQWP05M-700 by Thorlabs, Inc.). The induced polarization states are then normally incident onto the polarimetric imager. Images were taken with sufficient exposure time to ensure a high enough signal-to-noise ratio (SNR). The instrument matrix of each super-pixel was then calculated in MATLAB according to the transmitted intensities of each metasurface filter.

*Full Stokes polarization detection measurement.* Light coming from a High-Intensity Fiber coupled Halogen lamp light source (Thorlabs OSL2) is firstly collimated using a parabolic mirror (Thorlabs MPD129-P01), and the iris is applied to control the beam divergence angle. The bandpass filter (red: FBH650-40 green: FBH500-40) is applied for wavelength selection. Lens with F=30mm (AC254-030-AB) and F=100mm (AC254-100-AB) are applied for beam expansion. The final beam divergence is controlled to be 0.5 degrees with a spot size of 9mm in diameter. Arbitrary polarization states were generated using a broadband linear polarizer (WP25M-UB by Thorlabs, Inc.) and super achromatic QWP (SAQWP05M-700 by Thorlabs, Inc.) A list of Stokes parameters was first designed, then each polarization state was normally incident onto the polarimetric imager. These polarization states were firstly measured by the rotation of a linear analyzer (LPIREA100-C); the transmitted intensity was then fitted to obtain the polarization states. Afterward, the polarization states were captured by the polarimetric imager at a normal incidence angle. The images were then transferred onto a computer to calculate polarization states according to the transmitted intensities and the calibrated instrument matrix of each metasurface filter. The code for extracting polarization states for all super-pixels as well as Stokes parameter measurement, is performed in a MATLAB environment.

*Full Stokes polarization imaging.* A camera zoom lens is applied for imaging purposes. A color filter is attached in front of the lens. The field of view of the camera lens applied is ~ ±20 degrees for imaging demonstration.

**Acknowledgments:** These devices were fabricated in the Center for Solid State Electronics Research (CSSER) at Arizona State University.

**Funding:** This work was supported in part by NSF under Grant No. 2048230 and 1809997, and DOE under Grant No. DE-EE0008999. Device fabrication and characterization in the Center for Solid State Electronics Research (CSSER) and LeRoy Eyring Center for Solid State Science (LE-CSSS) at Arizona State University was supported, in part, by NSF contract ECCS-1542160.

**Author contributions:** Y.Y., J. Z and C. W. conceived the idea, J.Z and A. B performed the theoretical analysis, J. Z., J. B, S. C and X. C performed the device fabrication, J.Z and J. B, performed the device characterization, J. Z performed the data analysis. J. Z and Y. Y wrote the manuscript. All authors analyzed the results and contributed to the manuscript.

**Competing interests:** The authors declare no competing interests.

**Data and materials availability:** All data needed to the conclusions in the paper are present in the paper and/or the Supplementary Materials.